\documentclass[prl,twocolumn,nofootinbib,nobibnotes,noeprint,preprintnumbers,floatfix]
{revtex4-2}
\usepackage{amssymb,amsmath,graphicx}
\usepackage[dvipsnames]{xcolor}
\usepackage[colorlinks,allcolors=NavyBlue]{hyperref}
\usepackage{standalone}
\usepackage{gensymb}
\usepackage[splitrule]{footmisc}
\usepackage{longtable}
\preprint{FERMILAB-PUB-23-XXX-T}

\begin{document}

\preprint{FERMILAB-PUB-23-462-T}

\title{The Dark Matter Discovery Potential of the \\ Advanced Particle-Astrophysics Telescope (APT)}

\author{Fei Xu$^{a,b}$}
\thanks{feixu@uchicago.edu, ORCID: 0000-0002-8850-7580}

\author{Dan Hooper$^{a,b,c}$}
\thanks{dhooper@fnal.gov, ORCID: 0000-0001-8837-4127}

\affiliation{$^a$University of Chicago, Department of Astronomy and Astrophysics, Chicago, Illinois 60637, USA}
\affiliation{$^b$University of Chicago, Kavli Institute for Cosmological Physics, Chicago, Illinois 60637, USA}
\affiliation{$^c$Fermi National Accelerator Laboratory, Theoretical Astrophysics Group, Batavia, Illinois 60510, USA}

\date{\today}

\begin{abstract}

Gamma-ray observations of Milky Way dwarf galaxies have been used to place stringent constraints on the dark matter's annihilation cross section. In this paper, we evaluate the sensitivity of the proposed Advanced Particle-astrophysics Telescope (APT) to dark matter in these systems, finding that such an instrument would be capable of constraining thermal relics with masses as large as $m_X\sim 600 \, {\rm GeV}$. Furthermore, in dark matter scenarios motivated by the observed Galactic Center Gamma-Ray Excess, we predict that APT would detect several dwarf galaxies with high-significance. Such observations could be used to test the predicted proportionality between the gamma-ray fluxes and $J$-factors of individual dwarf galaxies, providing us with an unambiguous test of the origin of the Galactic Center Excess.

\end{abstract}

\maketitle

\section{Introduction}

For decades, the most widely studied dark matter candidates have been thermal relics of the early universe~\cite{Bertone:2016nfn}. Such particles were motivated by the realization that, if they had weak-scale masses and couplings, they would freeze out of equilibrium with an abundance that is approximately equal to that of the measured dark matter density -- a fact frequently referred to as the ``WIMP miracle.'' Many WIMP models, however, have since been ruled out by the null results of direct~\cite{LZ:2022ufs,XENON:2023sxq} and indirect~\cite{Fermi-LAT:2016uux,Calore:2022stf,Bergstrom:2013jra,Planck:2018vyg} searches. In light of this progress, some have argued that the WIMP paradigm is now disfavored. Simultaneously, it is not difficult to identify WIMP models which continue to be consistent with all existing constraints (for a discussion, see Ref.~\cite{Bertone:2018krk}).

The WIMP paradigm has been bolstered in recent years by the signal known as the Galactic Center Gamma-Ray Excess, as identified within the data collected by the Fermi Gamma-Ray Space Telescope~\cite{Cholis:2021rpp,DiMauro:2021raz} (for early work, see Refs.~\cite{Goodenough:2009gk,Hooper:2010mq,Hooper:2011ti,Abazajian:2012pn,Hooper:2013rwa,Gordon:2013vta,Daylan:2014rsa,Calore:2014xka,Fermi-LAT:2015sau}). The spectrum and angular distribution of this excess are each consistent with those predicted from annihilating dark matter. In particular, the measured characteristics of this signal are well fit by dark matter particles with a mass of $m_X \sim 35-60 \, {\rm GeV}$ and an annihilation cross section of $\langle \sigma v \rangle \sim (1-3) \times 10^{-26} \, {\rm cm}^3/{\rm s}$ (for the case of annihilations to $b\bar{b}$). This agrees remarkably well with the annihilation cross section expected of a thermal relic, $\langle \sigma v \rangle \approx 2.2 \times 10^{-26} \, {\rm cm}^3/{
\rm s}$ (see, for example, Ref.~\cite{PhysRevD.86.023506}).

If the Galactic Center Gamma-Ray Excess is generated by annihilating dark matter, an analogous signal should be produced from dwarf galaxies (for a review, see Ref.~\cite{Strigari:2018utn}). Gamma-ray observations of the Milky Way's dwarf galaxy population thus have the potential to confirm or rule out dark matter interpretations of the Galactic Center excess and, more broadly, to test the WIMP paradigm itself. The most recent such analysis by the Fermi Collaboration studied six years of data from the directions of 45 dwarf galaxies (and dwarf galaxy candidates), allowing them to rule out dark matter annihilation cross sections greater than $\langle \sigma v \rangle \sim 2.2 \times 10^{-26} \, {\rm cm}^3/{\rm s}$ for masses up to $m_X \sim 50 \, {\rm GeV}$ (again, for the case of annihilations to $b\bar{b}$)~\cite{Fermi-LAT:2016uux}. More recently, Di Mauro {\it et al.}~used 14 years of Fermi data to study a sample of 22 dwarf galaxies, producing similar constraints~\cite{DiMauro:2022hue}. Perhaps more interesting, these analyses also identified what could be hints of a dark matter annihilation signal. In particular, Ref.~\cite{DiMauro:2022hue} reports the presence of gamma-ray excesses from 
%that the Fermi data favors the presence of dark matter annihilation products from 
the dwarf galaxies Reticulum II, Sculptor, and Tucana II at a level of ${\rm TS} \approx 11$, 9, and 6, respectively (for the case of $m_X=50 \, {\rm GeV}$), where TS is the log-likelihood test statistic. Overall, this stacked analysis favors the presence of annihilating dark matter over the null hypothesis at a level of ${\rm TS}\approx 11$, corresponding to a local significance of 3.0$\sigma$~\cite{DiMauro:2022hue} (see also, Refs.~\cite{Fermi-LAT:2015ycq,Hooper:2015ula,Bhattacharjee:2018xem}).

Whereas gamma-ray studies of the Galactic Center are currently limited by systematic uncertainties associated with bright and poorly understood backgrounds, searches for gamma rays from dwarf galaxies are statistically limited. Future observations of the Milky Way's dwarf galaxy population with a large-acceptance, space-based gamma-ray telescope could thus significantly increase our sensitivity to dark matter and clarify the origin of the Galactic Center Gamma-Ray Excess.

\begin{table*}[] \small
\begin{tabular}{lc cc cc c cc cccc}
\hline
Dwarf Galaxy & Distance (kpc) & $\log_{10} J (0.5\degree)$ & $l$ ($\degree$) & $b$ ($\degree$)& Included in Fig.~1 \\
\hline
Canes Venatici I & 210.0 $\pm$ 6.0 & $17.42_{-0.15}^{+0.17}$ & 74.30 & 79.83 & No\\
Carina & 105.6 $\pm$ 5.4 & $17.83_{-0.09}^{+0.10}$ & 260.11	& -22.22 & Yes\\
Draco & 76.0 $\pm$ 6.0 & $18.83_{-0.12}^{+0.12}$ & 86.37 & 34.71& Yes\\
Fornax & 147.0 $\pm$ 9.0 & $18.09_{-0.10}^{+0.10}$ & 237.24	& -65.67 & Yes\\% include in Fermi not in APT
Leo I & 258.2 $\pm$ 9.5 & $17.64_{-0.12}^{+0.14}$ & 225.98	& 49.11 & No \\
Leo II & 233.0 $\pm$ 15.0 & $17.76_{-0.18}^{+0.22}$ & 220.16 & 67.23 & Yes\\
Sculptor & 83.9 $\pm$ 1.5 & $18.58_{-0.05}^{+0.05}$ & 287.70 & -83.15 & Yes\\
Sextans & 92.5 $\pm$ 2.2 & $17.73_{-0.12}^{+0.13}$ & 243.50	& 42.27 & Yes\\
Ursa Minor & 76.0 $\pm$ 4.0 & $18.75_{-0.12}^{+0.12}$ & 104.98 & 44.81 & Yes\\
Aquarius II & 107.9 $\pm$ 3.3 & $18.27_{-0.58}^{+0.66}$ & 55.11	& -53.01 & No\\
Boötes I & 66.0 $\pm$ 3.0 & $18.17_{-0.29}^{+0.31}$ & 358.10 & 69.64 & Yes\\
Canes Venatici II & 160.0 $\pm$ 7.0 & $17.82_{-0.47}^{+0.47}$ & 113.57 & 82.70 & Yes\\
Carina II & 37.4 $\pm$ 0.4 &  $18.25_{-0.54}^{+0.55}$ & 269.98 & -17.14 & No\\
Coma Berenices & 42.0 $\pm$ 1.5 & $19.00_{-0.35}^{+0.36}$ & 241.86 & 83.61 & Yes\\
Draco II & 20.0 $\pm$ 3.0 &  $18.93_{-1.70}^{+1.39}$ & 98.32 & 42.88 & No\\
Grus I & 120.2 $\pm$ 11.1 & $16.88_{-1.66}^{+1.51}$ & 338.65 & -58.24 & No\\
Hercules & 132.0 $\pm$ 6.0 & $17.37_{-0.53}^{+0.53}$ & 28.73 & 36.86 & Yes\\
Horologium I B & 87.0 $\pm$ 8.0 & $18.79_{-0.86}^{+0.90}$ & 271.38 & -54.74 & No\\
%Horologium I K & 79.0 $\pm$ 7.0 & $19.27_{-0.71}^{+0.77}$ & 271.38 & -54.74& No\\
Leo IV & 154.0 $\pm$ 5.0 & $16.40_{-1.15}^{+1.01}$ & 265.46	& 56.51 & Yes\\
Leo V & 173.0 $\pm$ 5.0 & $17.65_{-1.03}^{+0.91}$ & 261.86 & 58.53 & No\\
Pegasus III & 215.0 $\pm$ 12.0 & $18.30_{-0.97}^{+0.89}$ & 69.85 & -41.83 & No\\
Pisces II & 183.0 $\pm$ 15.0 & $17.30_{-1.09}^{+1.00}$ & 79.21 & -47.11 & No\\
Reticulum II B & 32.0 $\pm$ 2.0 & $18.88_{-0.37}^{+0.39}$ & 266.30 & -49.74 & No\\
%Reticulum II K & 30.0 $\pm$ 2.0 & $18.96_{-0.37}^{+0.38}$ & 266.30	& -49.74 &  No\\
Segue 1 & 23.0 $\pm$ 2.0 & $19.12_{-0.58}^{+0.49}$ & 220.48	& 50.41 & Yes\\
%Segue 2 & 36.6 $\pm$ 2.45 & $$\\
%Triangulum II & 30.0 $\pm$ 2.0 & $$\\
%Tucana II K & 57.5 $\pm$ 5.3 & $18.84_{-0.50}^{+0.55}$ & 328.09 & -52.32 & No\\
Tucana II B & 57.5 $\pm$ 5.3 & $19.02_{-0.53}^{+0.58}$ & 328.09	& -52.32 & No\\
%Tucana III & 25.0 $\pm$ 2.0 & $$\\
Ursa Major I & 97.3 $\pm$ 5.85 & $18.26_{-0.27}^{+0.29}$ & 159.36 & 54.43 & No\\
Ursa Major II & 34.7 $\pm$ 2.1 & $19.44_{-0.39}^{+0.41}$ & 152.46 & 37.44 & Yes\\
Willman 1 & 38.0 $\pm$ 7.0 & $19.53_{-0.50}^{+0.50}$ & 158.57 & 56.78 & Yes\\
Cetus & 780.0 $\pm$ 40.0 & $16.28_{-0.19}^{+0.20}$ & 156.47	& -78.53 & No\\
Eridanus II & 366.0 $\pm$ 17.0 & $17.28_{-0.31}^{+0.34}$ & 249.78 & -51.64 & No\\
Leo T & 407.0 $\pm$ 38.0 & $17.49_{-0.45}^{+0.49}$ & 214.85	& 43.66 & No\\
\hline
\end{tabular} 
\caption{The dwarf galaxies considered in this analysis. Note that we have excluded Fornax from our main analysis (due to it containing globular clusters~\cite{2021ApJ...923...77P}). The $J$-factor estimate are from Ref.~\cite{Pace:2018tin} and are given in units of GeV$^2$/cm$^5$.}
\label{tab:galcat}
\end{table*}

In this paper, we evaluate the sensitivity of the proposed Advanced Particle-astrophysics Telescope (APT) \cite{Alnussirat:2021tlo, APT:2021lhj} to dark matter annihilating in the dwarf galaxies of the Milky Way.\footnote{The proposed APT~\cite{Alnussirat:2021tlo, APT:2021lhj} is planned to follow the ADAPT (Antarctic Demonstrator for the Advanced Particle-astrophysics Telescope) mission, which is scheduled for a 30 day, sub-orbital flight in 2025. For more details, see \url{https://adapt.physics.wustl.edu/}.} To this end, we perform a stacked analysis of simulated data from the directions of 30 Milky Way dwarf galaxies, assessing the constraints on annihilating dark matter that could be achieved by such an instrument. We then estimate the projected sensitivity of APT to dark matter in a scenario motivated by the Galactic Center Gamma-Ray Excess. If the Galactic Center Excess is generated by annihilating dark matter, we find that APT will be able to detect gamma-ray signals from several dwarf galaxies at high significance. We conclude that such a telescope would be able to definitively confirm or rule out dark matter interpretations of the long-standing Galactic Center Gamma-Ray Excess.

\section{Dark Matter Annihilation in Milky Way Dwarf Galaxies}

%%%%%%%%%%%%%%%%%%%%%%%%%%%

%%%%%%%%%%%%%%%%%%%%%%%%%%%%

\begin{figure*}[t]
 \centering
  \includegraphics[width=16.0cm]{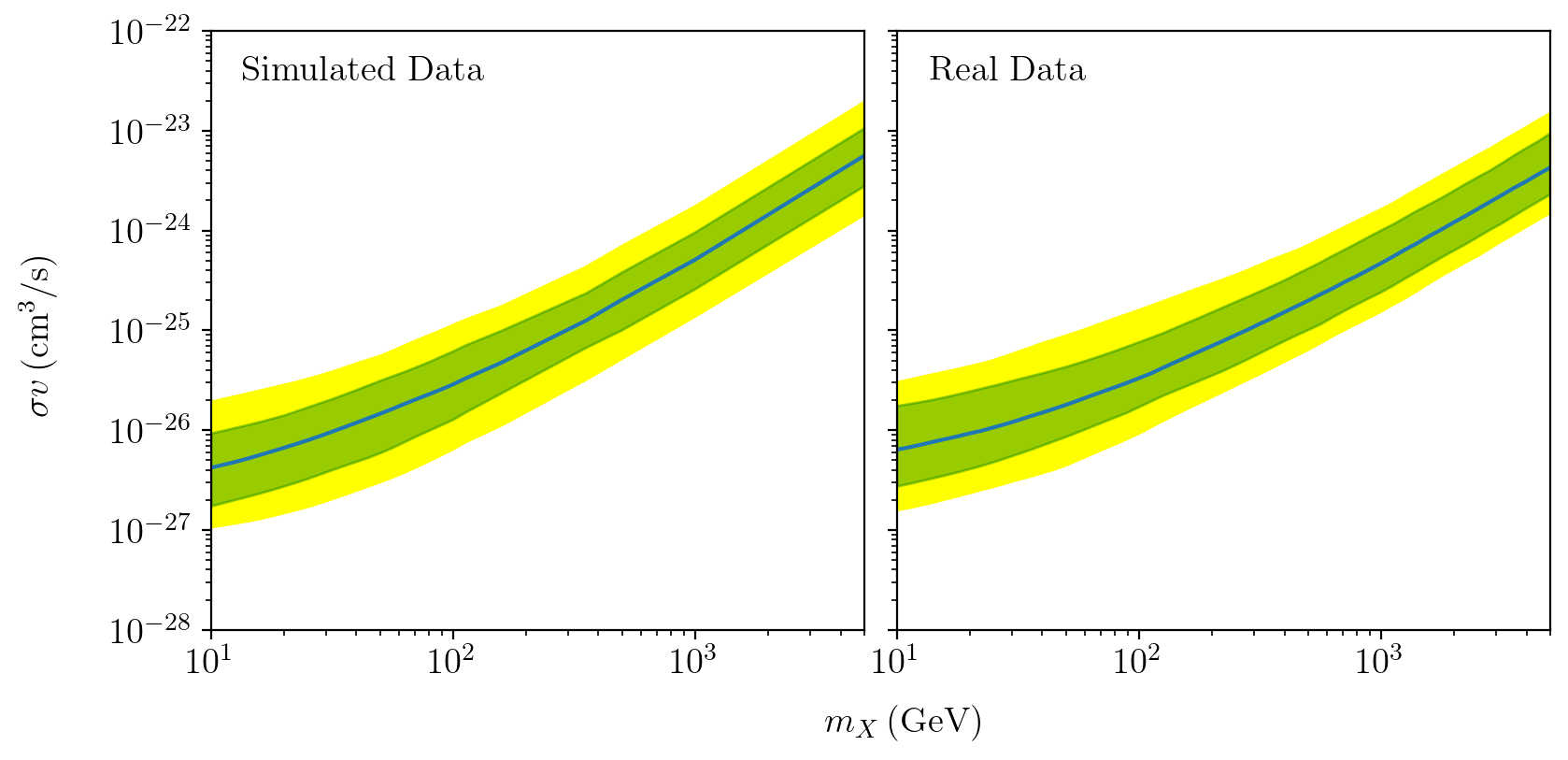}
  \caption{A comparison of the constraints on the dark matter annihilation cross section (to $b\bar{b}$) attained from four years of simulated (left) or real (right) Fermi data, from the directions of 15 dwarf galaxies. The solid lines and the surrounding green and yellow bands denote the median constraint and the range of constraints attained in 68\% and 95\% of the simulated realizations, respectively. In the right frame, we show the results from Ref.~\cite{Fermi-LAT:2013sme}, which were derived from four years of real Fermi data. The similarity between the simulated and real constraints demonstrates that our model provides an adequate description of the relevant backgrounds.}
  \label{fig:compare_Fermi}
\end{figure*}

The gamma-ray signal from annihilating dark matter can be calculated as follows:
\begin{align}
\frac{dN_{\gamma}}{dE_{\gamma}} &= \frac{\langle \sigma v \rangle}{8 \pi m_X^2} \, \frac{dN_{\gamma}}{dE_{\gamma}}\bigg|_{\rm ann}  \int_{\Delta \Omega} \int_{\rm los} \rho_X^2 \, dl \, d\Omega  \\
&= \frac{\langle \sigma v \rangle}{8 \pi m_X^2}  \,  \frac{dN_{\gamma}}{dE_{\gamma}}\bigg|_{\rm ann} \,J(\Delta \Omega), \nonumber
\end{align}
where $\langle \sigma v \rangle$ is the thermally-averaged dark matter annihilation cross section, $m_X$ is the mass of the dark matter particle, and $dN_{\gamma}/dE_{\gamma}|_{\rm ann}$ is the spectrum of gamma rays produced per annihilation. The $J$-factor, $J(\Delta \Omega)$, is defined as the square of the dark matter density, $\rho_X$, integrated over a solid angle, $\Delta \Omega$, and along the line-of-sight, $l$. The dark matter distributions of the Milky Way dwarf galaxies (and their corresponding $J$-factors) are constrained by spectroscopic measurements of stellar velocities. In our main analysis, we have used the $J$-factors provided by Pace and Strigari in Ref.~\cite{Pace:2018tin}.

Dwarf galaxies are attractive targets for dark matter searches due to their low astrophysical backgrounds. As gamma-ray telescopes become larger and more capable of detecting fainter sources, however, even modest backgrounds will become increasingly important. In our analysis, we have adopted a background model that consists of three components: 1) the gamma rays associated with unresolved point sources, 2) the background of isotropically distributed gamma rays, and 3) the diffuse emission associated with cosmic ray interactions in the interstellar medium. Note that these backgrounds do not originate from the dwarf galaxies themselves, but rather are associated with emission that is coincidentally produced along the lines-of-sight to these systems. Although millisecond pulsars in dwarf galaxies could potentially represent another background for dark matter searches, the gamma-ray fluxes arising from such objects are not expected to significantly impact such efforts~\cite{Winter:2016wmy}. Possible exceptions are the Fornax and Sagittarius dwarf galaxies, which are known to contain globular clusters~\cite{2021ApJ...923...77P,Evans:2022zno}, making it more likely that they harbor a significant population of millisecond pulsars. To be conservative, we have not included Fornax or Sagittarius in our main analysis.

Far away from the Galactic Plane, the gamma-ray emission from unresolved point sources is dominated by blazars, and we take the source count distribution of this population to follow the triply-broken power-law model described in Ref.~\cite{Marcotulli:2020fpm}. For each simulated observation of a dwarf galaxy, we draw from this distribution (up to sources as bright as $S \sim 10^{-6} \, {\rm ph} \,{\rm cm}^{-2} \, {\rm s}^{-1}$ and extrapolated down to $S \sim 10^{-13}$, where $S$ is the photon flux integrated above 0.1 GeV) to determine whether any such sources are present in that direction and, if so, their fluxes. We take the spectral shape of each blazar to be the same as that of the measured extragalactic gamma-ray background~\cite{Fermi-LAT:2014ryh}. Note this component of the gamma-ray background is fundamentally non-Poissonian in nature. The integral of this distribution constitutes approximately 60\% of the total extragalactic gamma-ray background. We take the remaining 40\% of this background to be isotropic, arising from diffuse mechanisms or from sources that produce no more than one photon in the data sets we will consider here. For the Galactic diffuse emission, we adopt the model $glliemv05.fit$, as provided in Ref.~\cite{Fermi-LAT:2013sme}. Unlike contributions from unresolved blazars and the isotropic background, the spectrum and intensity of the Galactic diffuse emission depends on the location of a given dwarf galaxy on the sky.

\begin{figure*}[t]
 \centering
  \includegraphics[width=16.0cm]{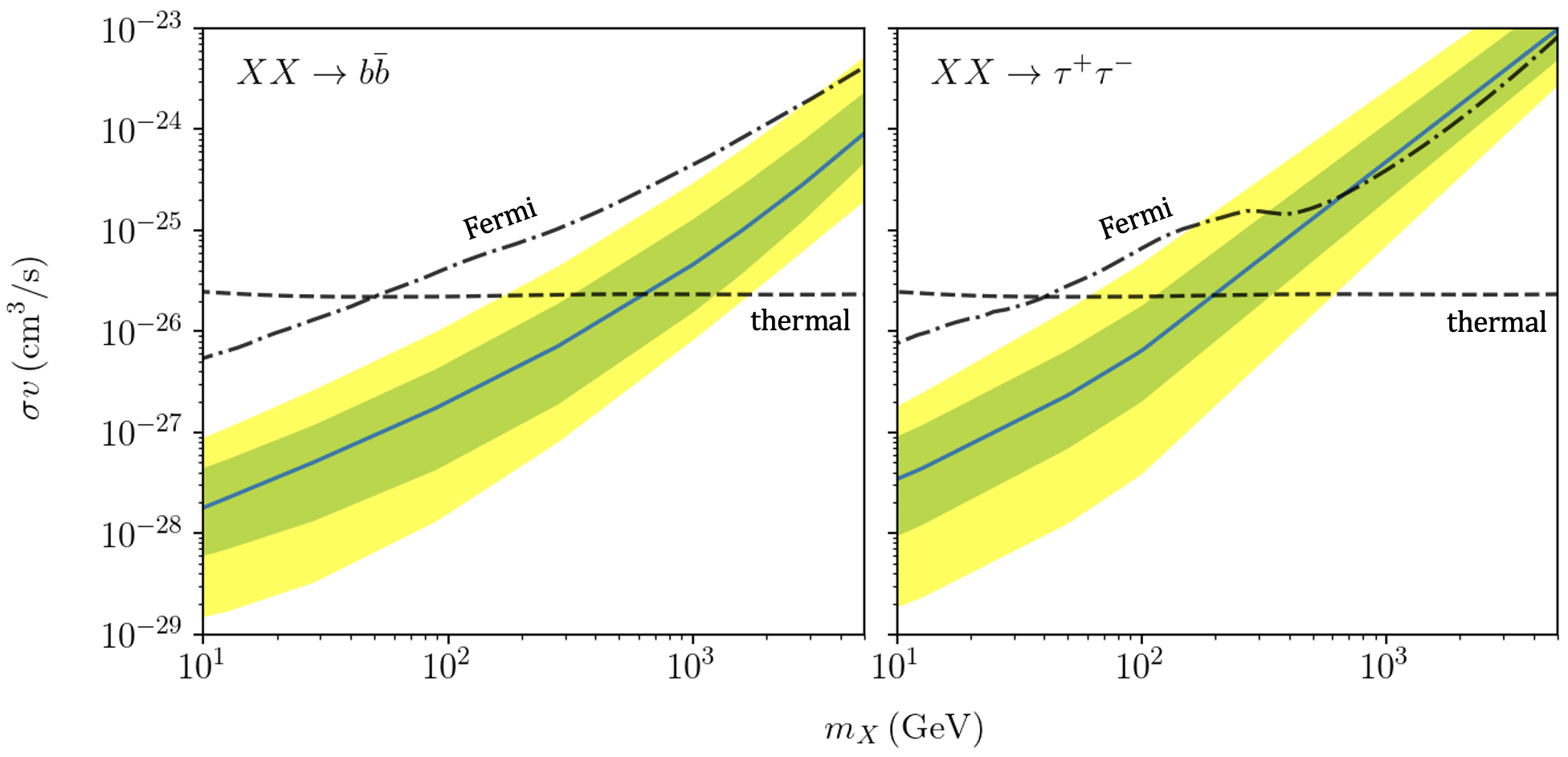}
  \caption{The projected constraints on the dark matter annihilation cross section (to $b\bar{b}$ or $\tau^+ \tau^-$) for 10 years of APT data from the directions of 30 Milky Way dwarf galaxies (see Table~\ref{tab:galcat}). The solid lines and the surrounding green and yellow bands denote the median constraint and the range of constraints attained in 68\% and 95\% of the simulated realizations, respectively. The dashed curve is the annihilation cross section predicted for a dark matter candidate that is a (velocity-independent) thermal relic~\cite{PhysRevD.86.023506}, while the dot-dashed line is the current constraint from Fermi data, as presented in Ref.~\cite{2022PhRvD.106l3032D}.}
  \label{fig:constraints}
\end{figure*}

\begin{figure*}[h]
  \hspace{-0.3cm}
  \includegraphics[width=16cm]{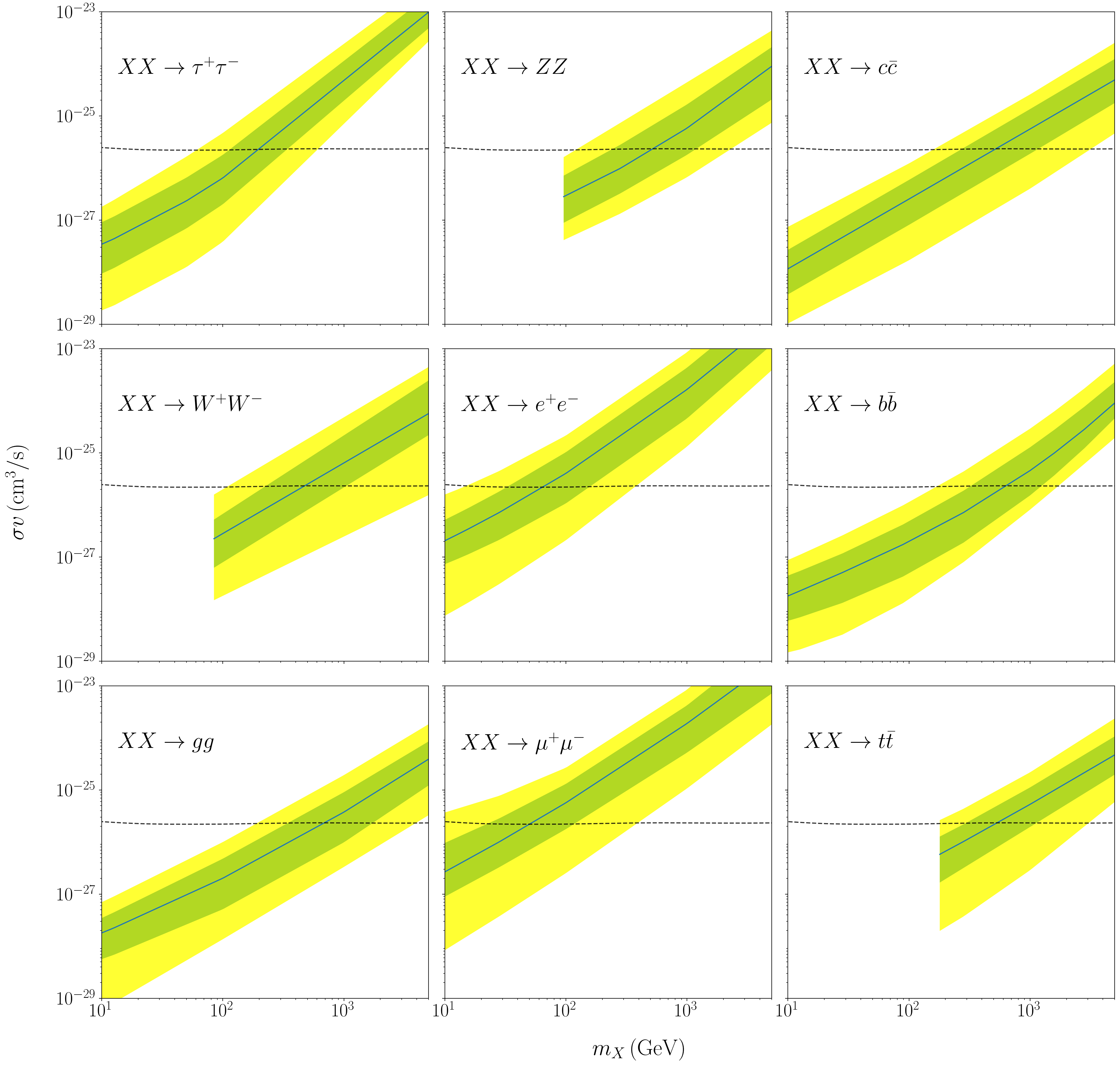} 
  \caption{As in Fig.~\ref{fig:constraints} but for other annihilation channels. These projected constraints are based on 10 years of simulated APT data from the directions of 30 Milky Way dwarf galaxies (see Table~\ref{tab:galcat}). The solid lines and the surrounding green and yellow bands denote the median constraint and the range of constraints attained in 68\% and 95\% of the simulated realizations, respectively. The dashed curve is the annihilation cross section predicted for a dark matter candidate that is a (velocity-independent) thermal relic~\cite{PhysRevD.86.023506}.}
  \label{fig:all_channel}
  \vspace{0cm}
\end{figure*}

Once we have determined the total gamma-ray flux from the direction of a given dwarf galaxy, we integrate over each energy bin and multiply by the instrumental exposure (using the acceptance evaluated at the average energy of that bin) to determine the mean number of photons that are observed in each energy bin and from within $0.5^{\circ}$ of the dwarf galaxy's center. We then draw from a Poisson distribution in each bin to determine the number of photons that are observed in that realization, and use this information to calculate the likelihood of attaining the simulated data as a function of $\langle \sigma v\rangle$ (for a given value of the dark matter mass and annihilation channel). Furthermore, for each simulated dwarf, we draw the value of $\log_{10}(J)$ from a Gaussian distribution with a central value and width equal to the quoted best-fit value and uncertainty~\cite{Pace:2018tin}. In evaluating the stacked likelihood, we follow the approach described in Ref.~\cite{Fermi-LAT:2013sme}. We repeat this procedure 1000 times for each dark matter mass, cross section, and channel in order to attain the resulting statistical distribution.

To assess the adequacy of our simulation and background model, we simulated four years of Fermi data from the directions of 15 dwarf galaxies and compared the resulting constraints to those attained by the Fermi Collaboration from four years of real data~\cite{Fermi-LAT:2013sme}. In performing this comparison, we adopted the same $J$-factors (and their uncertainties) as in Ref.~\cite{Fermi-LAT:2013sme}, and used the energy-dependent acceptance as given in Ref.~\cite{Fermi-LAT:2021wbg}. For each realization, we calcross sectionculated the 95\% confidence-level upper limit on the annihilation cross section, corresponding to a change in the log-likelihood (relative to $\langle \sigma v\rangle=0$) of $2 \Delta \ln \mathcal{L} = -3.84$. As can be seen in Fig.~\ref{fig:compare_Fermi}, our simulated constraints are very similar to those found using real Fermi data, demonstrating that our model provides a good description of the backgrounds relevant to such an analysis.

\section{The Projected Sensitivity of APT to Annihilating Dark Matter}

To assess the projected sensitivity of the Advanced Particle-astrophysics Telescope (APT)~\cite{Alnussirat:2021tlo, APT:2021lhj} to dark matter particles annihilating in Milky Way dwarf galaxies, we simulated 10 years of APT data, adopting an acceptance as given in Ref.~\cite{Alnussirat:2021tlo}, considering the 30 dwarf galaxies listed in Table~\ref{tab:galcat} (exempting Fornax), and using the $J$-factor determinations from Ref.~\cite{Pace:2018tin}.

We show the results of this exercise in Fig.~\ref{fig:constraints}, for the cases of annihilation to $b\bar{b}$ or $\tau^+ \tau^-$. Due to the much larger acceptance of APT, these projected constraints are significantly more stringent than those derived from Fermi data~\cite{Fermi-LAT:2016uux,DiMauro:2022hue}. For comparison, we also include in these frames the annihilation cross section for a dark matter candidate that is a (velocity-independent) thermal relic~\cite{PhysRevD.86.023506}, as well as the latest constraints derived from Fermi data~\cite{2022PhRvD.106l3032D}. The constraints projected for other annihilation channels are shown in Fig.~\ref{fig:all_channel}.

%for nine choices of annihilation channels. Due to the much larger acceptance of APT, these projected constraints are much more stringent than those derived from Fermi data (either as shown in Fig.~\ref{fig:compare_Fermi} or in more recent studies~\cite{Fermi-LAT:2016uux,DiMauro:2022hue}).

\section{Testing the Origin of the Galactic Center Gamma-Ray Excess}

\begin{figure}[t]
 \centering
  \includegraphics[width=8.7cm]{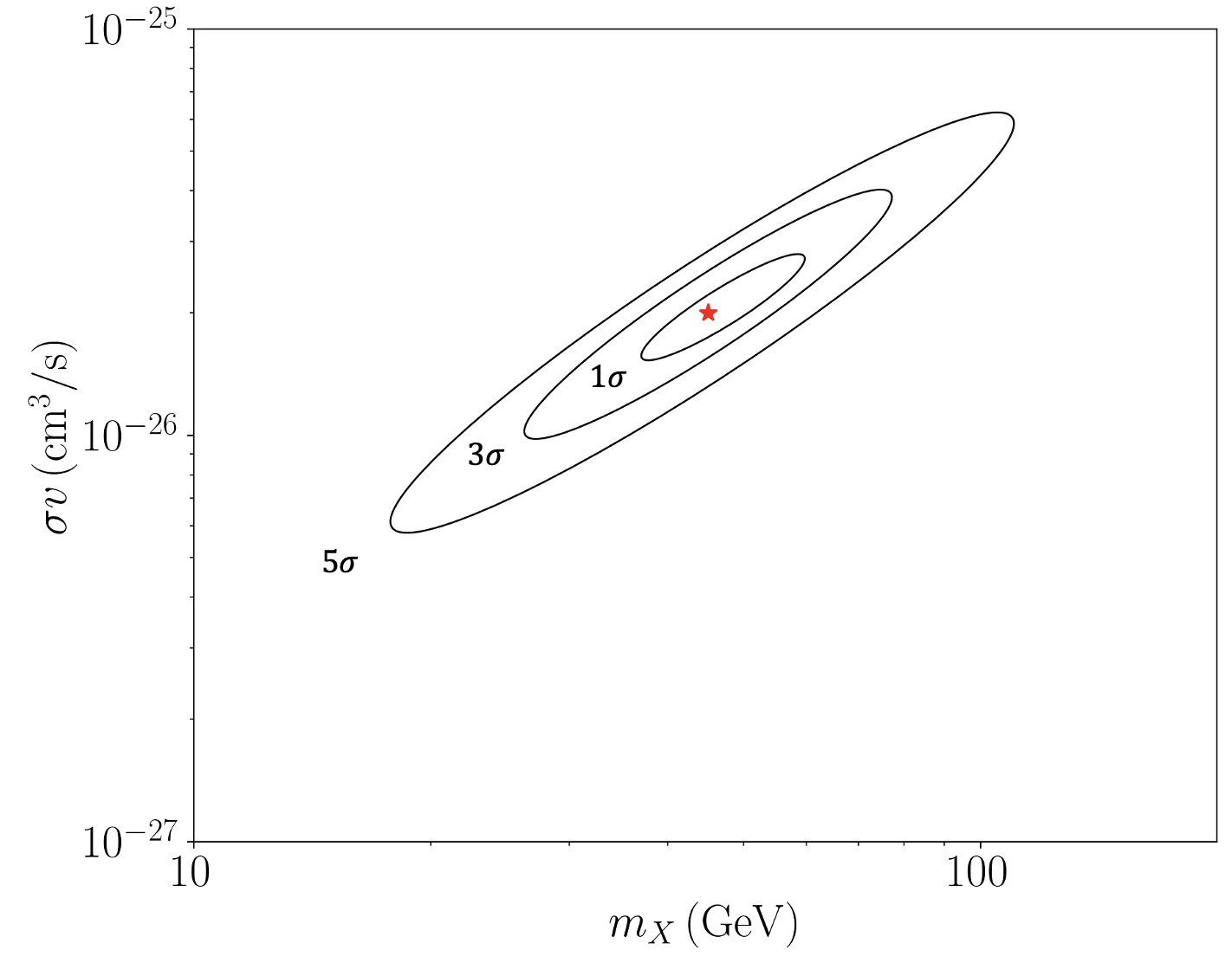}
  \caption{The projected ability of APT (with 10 years of data) to measure the dark matter mass and annihilation cross section in a scenario with $m_X=45 \, {\rm GeV}$ and $\langle \sigma v \rangle = 2 \times 10^{-26} \, {\rm cm}^3/{\rm s}$ (to $b\bar{b}$), as motivated by the Galactic Center Gamma-Ray Excess. The star and surrounding contours represent the best-fit value and the 1, 3 and 5$\sigma$ confidence intervals, respectively. In such a scenario, we project that APT could exclude the null hypothesis at a level of $2 \Delta \ln \mathcal{L} \approx -200$, corresponding to a significance of 14$\sigma$.}
  \label{fig:contour}
\end{figure}

\begin{figure}[t]
 \centering
  \includegraphics[width=9cm]{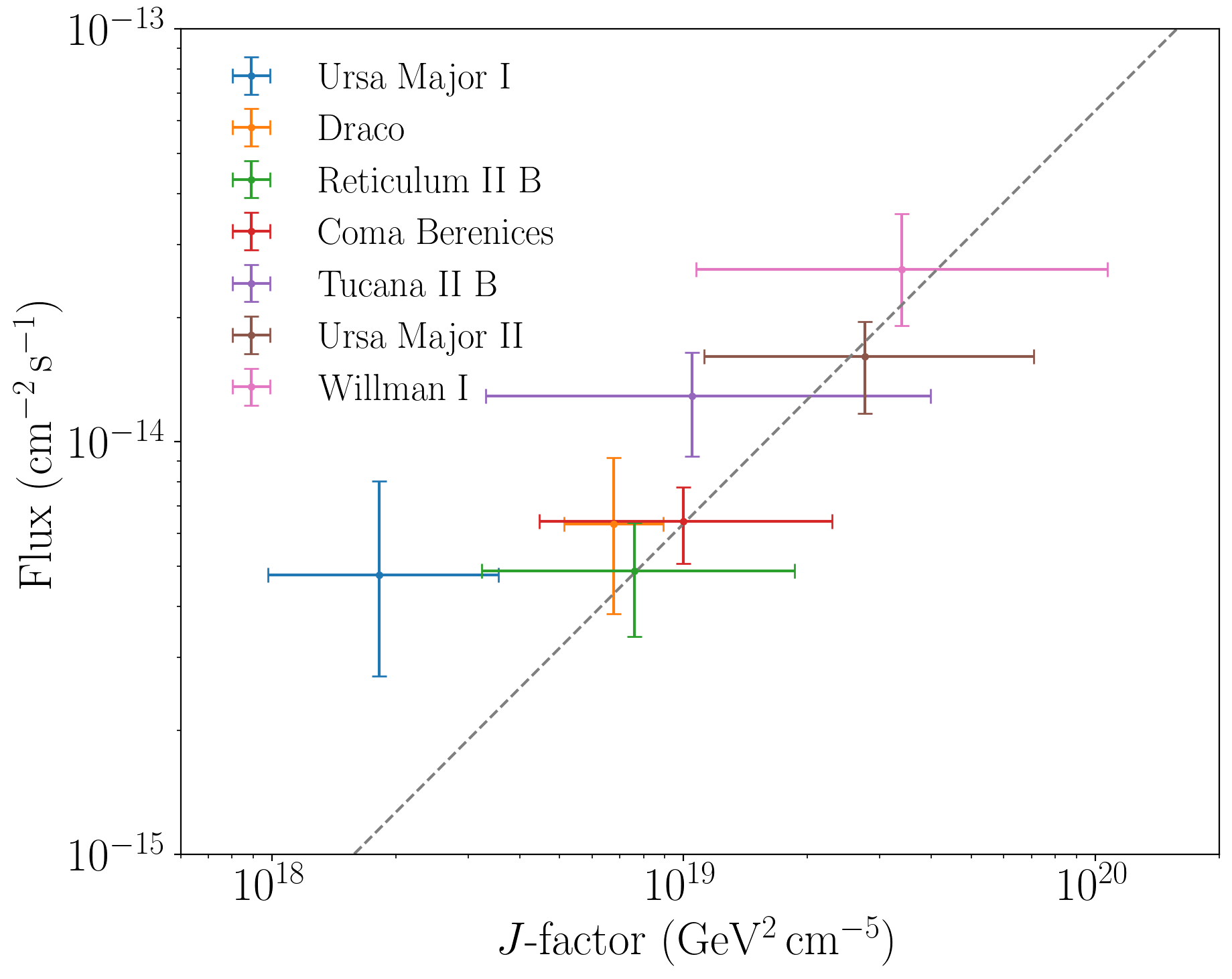}
  \caption{The projected ability of APT (with 10 years of data) to measure the gamma-ray fluxes (integrated above 0.1 GeV) from individual dwarf galaxies in a scenario with $m_X=45 \, {\rm GeV}$ and $\langle \sigma v \rangle = 2 \times 10^{-26} \, {\rm cm}^3/{\rm s}$ (to $b\bar{b}$), as motivated by the Galactic Center Gamma-Ray Excess. These fluxes are compared to the $J$-factors of the dwarfs, as integrated within a radius of $0.5^{\circ}$. These results were attained in a single (but representative) realization of our simulation, showing each dwarf that was detected with a significance of $2\sigma$ or higher. Such a data set would allow us to test whether the gamma-ray fluxes from dwarf galaxies are proportional to the corresponding $J$-factors, providing an unambiguous test of dark matter interpretations of the Galactic Center Gamma-Ray Excess.}
  \label{fig:errorbars}
\end{figure}

In the previous section, we carried out our simulations under the assumption that there is no signal from annihilating dark mater, and derived the constraints that could be attained by an instrument such as APT. It is possible, however, that such a signal could be found in the data, in particular in light of the long-standing Galactic Center Gamma-Ray Excess~\cite{Cholis:2021rpp,DiMauro:2021raz,Goodenough:2009gk,Hooper:2010mq,Hooper:2011ti,Abazajian:2012pn,Hooper:2013rwa,Gordon:2013vta,Daylan:2014rsa,Calore:2014xka,Fermi-LAT:2015sau}. Motivated by this excess, we consider in this section a scenario in which the Galactic Center excess is generated by annihilating dark matter, evaluating the sensitivity of APT to a dark matter candidate that is capable of generating this signal. 

For concreteness, we will consider a dark matter particle with a mass of $m_X=45 \, {\rm GeV}$ and that annihilates to $b\bar{b}$ with a cross section of $\langle \sigma v \rangle = 2 \times 10^{-26} \, {\rm cm}^3/{\rm s}$. We again perform a simulation of 10 years of APT data from the directions of 30 dwarf galaxies, but calculate the change in the log-likelihood relative to the best fit value of $m_X$ and $\langle \sigma v \rangle$. The results of this exercise are shown in Fig.~\ref{fig:contour}. Relative to the best-fit parameter values, $\langle \sigma v \rangle =0$ is disfavored in this scenario at a level of $2 \Delta \ln \mathcal{L} \approx -200$, ruling out the null hypothesis with a significance of approximately 14$\sigma$.

The gamma-ray signal from dark matter annihilating in a given dwarf galaxy is proportional to that galaxy's $J$-factor, providing us with a powerful way to distinguish dark matter annihilation products from astrophysical backgrounds (which would not be expected to scale with $J$). In a scenario with $m_X=45 \, {\rm GeV}$ and $\langle \sigma v \rangle = 2 \times 10^{-26} \, {\rm cm}^3/{\rm s}$ (to $b\bar{b}$), we find (in our median simulation) that APT will detect 7 dwarf galaxies at $>2\sigma$ significance, 4 at $>3\sigma$, and 3 at $>5\sigma$. This data would allow us to test the whether the the gamma-ray fluxes from these galaxies are, in fact, proportional to the corresponding $J$-factors. In Fig.~\ref{fig:errorbars}, we show the results of a representative realization of our simulation, showing the gamma-ray fluxes from each of the 7 dwarfs that were detected with greater than $2\sigma$ significance.

\section{Summary and Conclusions}
 
Gamma-ray observations of dwarf galaxies can be used to place stringent constraints on annihilating dark matter. At present, such searches are statistically limited, and thus would significantly benefit from experiments capable of detecting more gamma-ray photons from such systems. In this context, we have evaluated in this paper the sensitivity of a future, large-acceptance, space-based gamma-ray telescope, focusing on the case of the proposed Advanced Particle-astrophysics Telescope (APT). 

As shown in Figs.~\ref{fig:constraints} and~\ref{fig:all_channel}, we project that an APT-like telescope would be very sensitive to annihilating dark matter particles, probing annihilating cross sections associated with thermal relics for masses up to $m_X \sim 600 \, {\rm GeV}$ (for the case of annihilation to $b\bar{b}$). In contrast, Fermi is currently only sensitive to such particles if they are lighter than $m_X\sim 50 \, {\rm GeV}$~\cite{Fermi-LAT:2016uux,2022PhRvD.106l3032D}.
  
If the Galactic Center Gamma-Ray Excess is generated by annihilating dark matter, the corresponding signal from dwarf galaxies would be unambiguously detected by APT. In such a scenario, we find that APT would detect 7 dwarf galaxies with a significance of at least 2$\sigma$, and 3 dwarfs with a significance of 5$\sigma$ or greater (in our median simulation). From such measurements, it could be established whether the gamma-ray fluxes from dwarf galaxies are proportional to the corresponding $J$-factors, providing us with a smoking gun signature of annihilating dark matter.

In our simulations, we have considered 30 known dwarf galaxies and have used currently available determinations of their $J$-factors. It is anticipated, however, that many new dwarf galaxies will be discovered in the Rubin/LSST era, increasing the sensitivity of gamma-ray searches for dark matter annihilation products~\cite{LSSTDarkMatterGroup:2019mwo,Fermi-LAT:2016afa,He:2013jza,Hargis:2014kaa}. Furthermore, spectroscopic measurements of stellar velocities in dwarf galaxies will continue to improve our ability to determine the $J$-factors of these systems, further improving the sensitivity of stacked dwarf analyses of gamma-ray data. For these reasons, the actual sensitivity of APT to annihilating dark matter could plausibly exceed the projections that we have presented in this study.

\bigskip
\bigskip

\begin{acknowledgments}

We would like to thank Alex Drlica-Wagner for helpful discussions. FX would further like to thank Rich Kron for his ongoing support. This work is supported by the Fermi Research Alliance, LLC under Contract No.~DE-AC02-07CH11359 with the U.S. Department of Energy.

\end{acknowledgments}

% For when including just prebuilt bbl file for PRL or ArXiv
%\input{APT-PRL-vfinal.bbl}
% For when building bilbiography

\bibliography{APT-PRL-vfinal}

\end{document}